\documentstyle[12pt,preprint,aps,floats]{revtex}

\tighten

\input epsf.tex

\newcommand{\postscript}[2]{\setlength{\epsfxsize}{#2\hsize}
   \centerline{\epsfbox{#1}} \vspace*{.1in}}

\newcommand{\tb}{\tan\beta}
\newcommand{\fb}{\text{fb}}
\newcommand{\ifb}{\text{fb}^{-1}}
\newcommand{\mev}{\text{MeV}}
\newcommand{\gev}{\text{GeV}}
\newcommand{\tev}{\text{TeV}}

\newcommand{\yr}{\text{yr}}

\newcommand{\ecm}{E_{\text{CM}}}

\newcommand{\emem}{e^-e^-}
\newcommand{\epem}{e^+e^-}
\newcommand{\ser}{\tilde{e}_R}
\newcommand{\sel}{\tilde{e}_L}
\newcommand{\mser}{m_{\tilde{e}_R}}
\newcommand{\msel}{m_{\tilde{e}_L}}
\newcommand{\mchi}{m_{\chi}}
\newcommand{\ltot}{L_{\rm tot}}

\newcommand{\eqref}[1]{Eq.~(\ref{#1})}
\newcommand{\bold}[1]{{\text{\normalsize\boldmath $#1$}}}

\begin{document}

\draft

\renewcommand{\thefootnote}{\fnsymbol{footnote}}
\setcounter{footnote}{0}

\preprint{
\noindent
\hfill
\begin{minipage}[t]{3in}
\begin{flushright}
MIT--CTP--3136\\
SLAC--PUB--8829\\
hep-ph/0105100
\end{flushright}
\end{minipage}
}

\title{ \vskip 0.5in 
Selectron Studies at $\bold{\emem}$ and $\bold{\epem}$ 
Colliders\footnote{Work supported by the U.~S.~Department of Energy,
contract DE--AC03--76SF00515 and in part by the U.~S.~Department of
Energy under cooperative research agreement DF--FC02--94ER40818.}  }

\author{
Jonathan L.~Feng$^a$\footnote{E-mail: jlf@mit.edu} and 
Michael E. Peskin$^b$\footnote{E-mail:
mpeskin@slac.stanford.edu}
\vskip 0.2in
}

\address{
  ${}^{a}$
  Center for Theoretical Physics, 
  Massachusetts Institute of Technology\\
  Cambridge, MA 02139 USA\\ 
\vskip 0.1in
  ${}^{b}$
  Stanford Linear Accelerator Center, 
  Stanford University\\
  Stanford, CA 94309 USA}

\maketitle

\begin{abstract}
Selectrons may be studied in both $\emem$ and $\epem$ collisions at
future linear colliders.  Relative to $\epem$, the $\emem$ mode
benefits from negligible backgrounds and $\beta$ threshold behavior
for identical selectron pair production, but suffers from luminosity
degradation and increased initial state radiation and beamstrahlung.
We include all of these effects and compare the potential for
selectron mass measurements in the two modes.  The virtues of the
$\emem$ collider far outweigh its disadvantages.  In particular, the
selectron mass may be measured to $100~\mev$ with a {\em total\/}
integrated luminosity of $1~\ifb$, while more than $100~\ifb$ is
required in $\epem$ collisions for similar precision.
\end{abstract}

%\vspace{1.3cm}
%\centerline{\sl Submitted to Phys.~Rev.~D}

%\vspace{1.3cm} 
%\centerline{\sl PACS numbers: 11.10.Hi, 11.30.Pb, 12.60.Jv, 14.80.Ly,
%04.65.+e}

\newpage

\renewcommand{\thefootnote}{\arabic{footnote}}
\setcounter{footnote}{0}

\section{Introduction}
\label{sec:introduction}

If new particles exist at the weak scale, linear colliders are likely
to play an important role in determining their properties and
illuminating their relationships to electroweak symmetry breaking.
This is especially true for supersymmetric particles.  In $\epem$
collisions, linear colliders produce superpartners democratically, and
the ability to specify the initial partons' energies and (in the case
of electrons) spins makes possible a rich program of highly
model-independent measurements~\cite{Murayama:1996ec}.

The flexibility of the linear collider program is further enhanced by
the possibility of $\emem$, $e^- \gamma$, and $\gamma\gamma$
collisions.  The $\emem$ possibility is a prerequisite for the
$e^-\gamma$ and $\gamma \gamma$ modes, as highly polarized beams are
required to produce high energy back-scattered photons.  The $\emem$
mode is also an inexpensive and technologically trivial extension, and
provides an ideal environment for studying beam polarization, certain
precision electroweak observables, and a variety of exotic new physics
possibilities.  Studies of these and other topics may be found in
Refs.~\cite{95,97,99}.

In the case of supersymmetry, electric charge and lepton number
conservation imply that, in simple models, only selectrons are readily
produced in $\emem$ mode~\cite{Keung:1983nq}.  However, these same
symmetries also eliminate many potential backgrounds to selectron
events.  In addition, the unique quantum numbers of the $\emem$
initial state imply that threshold cross sections for identical
selectron pair production are proportional to $\beta$, the velocity of
the produced selectrons.  They therefore rise much more sharply than
in $\epem$ collisions, where the threshold cross section is
proportional to $\beta^3$.  For these and other reasons to be
described below, the $\emem$ mode provides a promising environment for
studies of selectrons, and sleptons in
general~\cite{Feng:1998ud,Feng:2000zv}.  The potential of $\emem$
colliders for high precision studies of slepton
flavor~\cite{Arkani-Hamed:1996au} and CP
violation~\cite{Arkani-Hamed:1997km} and super-oblique
parameters~\cite{Cheng:1997sq,Cheng:1998sn} has been considered
previously.

Here we explore the potential of $\emem$ collisions for selectron
threshold mass measurements. Precise measurements of superparticle
masses are required to determine the parameters of the weak-scale
supersymmetric Lagrangian and, ultimately, the underlying theory at
shorter length scales~\cite{Blair:2001gy}.  Threshold scans have great
potential, but are sensitive to beam luminosity profiles.  We consider
realistic beam designs as recently implemented in the {\tt pandora}
simulation package~\cite{PANDORA}.  These include the effects of
initial state radiation (ISR), beamstrahlung, beam energy spread, and
the luminosity reduction appropriate for $\emem$ collisions.  While
all of these effects degrade the results, they are more than
compensated for by the intrinsic benefits of $\emem$ collisions.  We
show, in particular, that selectron mass measurements at the part per
mil level may be achieved with a total integrated luminosity of $\ltot
= 1~\ifb$.  In contrast, for the $\epem$ mode, we find that, even
ignoring possibly large backgrounds, similar precision requires well
over $100~\ifb$.  Our $\epem$ results are roughly similar to those of
previous studies~\cite{Martyn:1999tc,Martyn:1999xc}, although
differing quantitatively.  The $\emem$ mode therefore provides
incomparable opportunities for high precision selectron mass
measurements with very little investment of luminosity.

\section{Selectron Production and Decay}
\label{sec:selectron}

Selectron pair production at $\emem$ colliders takes place through the
processes shown in Fig.~\ref{fig:selectron}.  In general, each final
state selectron may be either $\ser^-$ or $\sel^-$, and all four
neutralinos $\chi^0_i$ are exchanged in the $t$-channel.  General
characteristics of this production mechanism are discussed in
Ref.~\cite{Peskin:1998jy}.

\begin{figure}[tbp]
\postscript{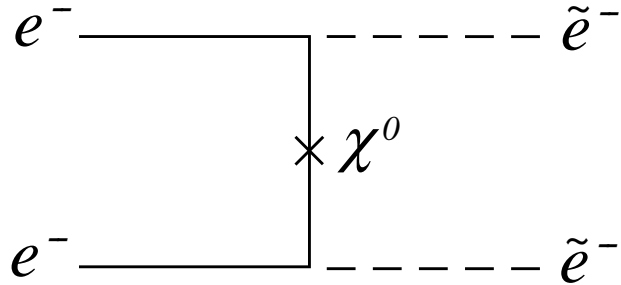}{0.59}
\caption{Selectron pair production $e^- e^- \to \tilde{e}^-
\tilde{e}^-$, mediated by $t$-channel neutralino exchange.}
\label{fig:selectron}
\end{figure}

In this study, we focus on the case of $\ser^-$ pair production.
Right-handed sleptons are neutral under both SU(3) and SU(2)
interactions.  In many supersymmetric models, they are therefore the
lightest scalars and so the most likely to be within kinematic reach
of linear colliders.  For $\emem \to \ser^- \ser^-$, only $t$-channel
Bino exchange contributes.  For simplicity, we will assume that the
lightest neutralino $\chi$ is a pure Bino with mass $\mchi= M_1$, and
we neglect the possibility of slepton flavor violation.  (These
assumptions may be tested experimentally at a linear collider, as we
discuss Sec.~\ref{sec:summary}.)  The production cross section for
$\emem \to \ser^- \ser^-$, then, depends on only two supersymmetry
parameters, $m_{\ser}$ and $M_1$.  The differential cross section is
\begin{equation}
\frac{d\sigma}{d\Omega} = \frac{\alpha^2 M_1^2}{2 \cos^4\theta_W}
\left( \frac{1}{t - M_1^2} + \frac{1}{u - M_1^2}\right)^2 \ ,
\label{crosssection}
\end{equation}
where the factor $M_1^2$ in the numerator arises from the Majorana
mass insertion required in the Bino propagator.  

In the reaction $\emem \to \ser^- \ser^-$, the initial state of two
right-handed electrons has angular momentum $L_z = 0$.  The selectrons
may then be produced in an $S$ wave state, and so at threshold the
cross section rises as $\beta$, the velocity of the outgoing
selectrons.  This contrasts sharply with the behavior of $\epem \to
\ser^+ \ser^-$.  In that reaction, the initial state is a right-handed
electron and a left-handed positron, and so has $L_z = 1$.  Selectrons
are then necessarily produced in a $P$ wave state, and the cross
section rises as $\beta^3$ at threshold.  This conclusion is based
solely on the properties of the initial and final states and is
independent of the relative importance of the $t$- and $s$-channel
contributions to $\epem \to \ser^+ \ser^-$.

Once produced, selectrons must decay.  In supergravity frameworks,
they typically decay via $\ser^- \to e^- \chi$.  For a Bino $\chi$,
the width is
\begin{equation}
\Gamma_{\ser} = \frac{\alpha \mser}{2 \cos^2 \theta_W} \left[ 1 -
\left(\frac{\mchi}{\mser}\right) ^2 \right] ^2 \ .
\label{width}
\end{equation}
In $R$-parity violating theories, selectrons may decay to three
standard model particles, and in theories with low-energy
supersymmetry breaking, selectrons may decay to gravitinos or through
three-body modes to staus.  If any of these is the dominant decay
mode, the selectron width is negligible for calculations of threshold
cross sections.

The threshold behavior of selectron production is shown in
Fig.~\ref{fig:epmcomp} for the case $(\mser, \mchi) = (150~\gev,
100~\gev)$ for both $\emem$ and $\epem$ modes and the beam designs
given in Table~\ref{table:I}.  We assume beam polarizations $P_{e^-} =
0.8$ and $P_{e^+}=0$, where
\begin{equation}
P \equiv \frac{N_R - N_L}{N_R + N_L} \ .
\end{equation}

\begin{figure}[btp]
\postscript{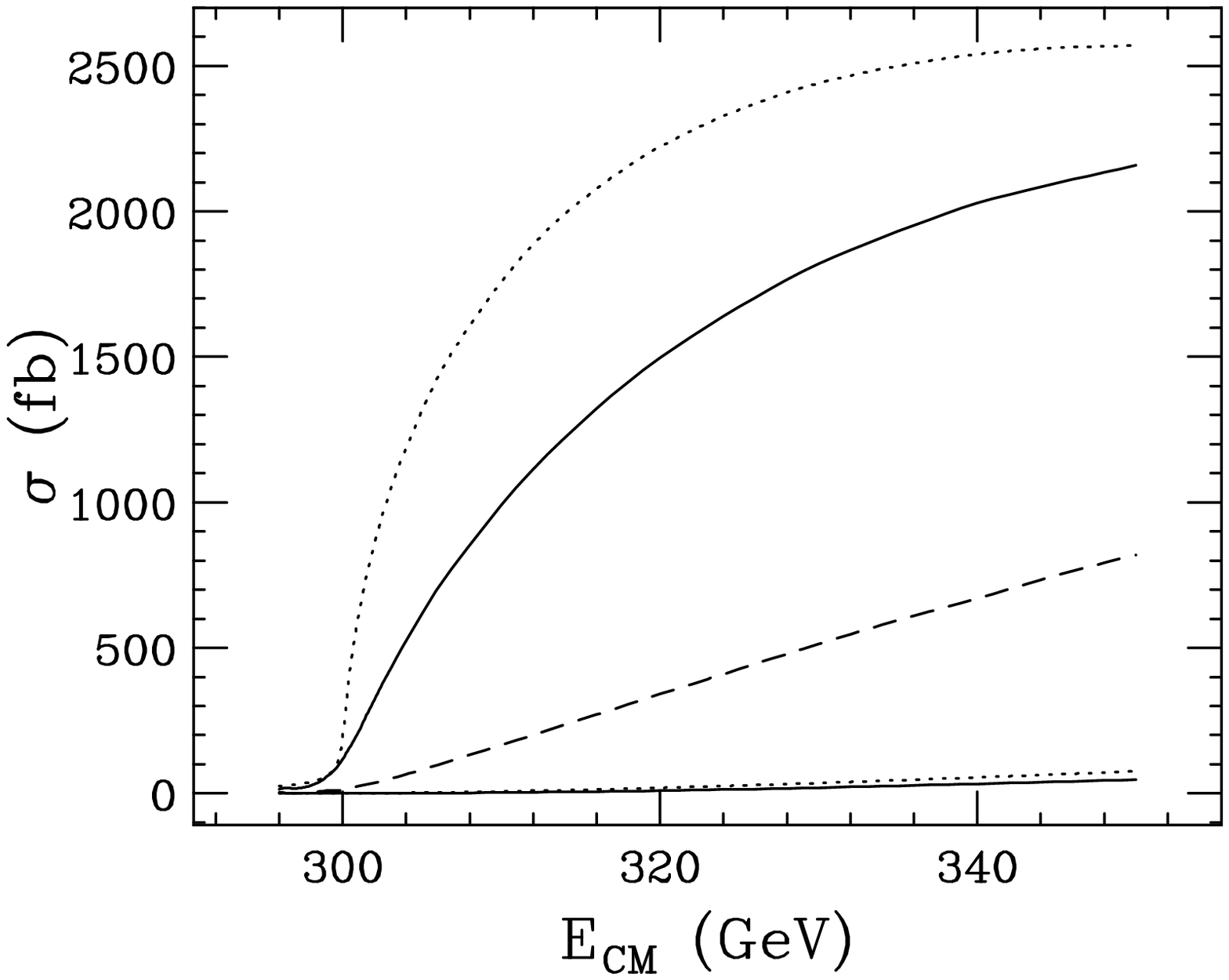}{0.59}
\caption
{Threshold behavior for $\sigma(\emem \to \ser^- \ser^-)$ (upper two
contours) and $\sigma(\epem \to \ser^+ \ser^-)$ (lower two contours)
for $(\mser, \mchi) = (150~\gev, 100~\gev)$.  In each pair, the dotted
curve neglects all beam effects, and the solid curve includes the
ISR/beamstrahlung and beam energy spread of the NLC500H flat beam
design. Results for $\emem$ EE500 round beams (dashed) are also shown.
Beam polarizations $P_{e^-} = 0.8$ and $P_{e^+}=0$ are assumed, and
the selectron width is included.  }
\label{fig:epmcomp}
\end{figure}

\begin{table}[btp]
\caption{Beam designs considered here.
\label{table:I}
}
\begin{tabular}{lccr}
& Type
& Mode
& \multicolumn{1}{c}{${\cal L}(300~\gev)$~($\ifb/\yr$)}  
\\ \hline
NLC500H~\cite{Thompson:2000ij} \rule[0mm]{0mm}{4mm} 
                     & flat  & $\emem$ & 78  \hspace*{.5in} \\
%NLC500A,B,C~\cite{Thompson:2000ij} 
%                     & flat  & $\emem$ & 28  \hspace*{.5in} \\
NLC500H~\cite{Thompson:2000ij} 
                     & flat  & $\epem$ & 240 \hspace*{.5in} \\
EE500~\cite{Zimmermann:1998bh}   
                     & round & $\emem$ & 44  \hspace*{.5in} \\
\end{tabular}
\vspace*{-.2in}
\end{table}

Our treatment of the beams includes the effects of initial state
radiation, beamstrahlung, and beam energy spread using approximate
parameterizations which treat the two beams independently.  For ISR,
we use the structure function prescription, with the form of the
structure function suggested by Skrzypek and
Jadach~\cite{Skrzypek:1991qs}.  For beamstrahlung, we generate the
spectrum from an approximate integral equation~\cite{myLC} which
improves upon the treatment of Yokoya and Chen~\cite{YC}.  This
procedure makes use of phenomenological parameterizations of beam
disruption at the collision due to Chen~\cite{Chen1,Chen2} for
$e^+e^-$ and to Chen and Thompson~\cite{Chen3,Thompson:2000ij} for
$e^-e^-$.  For beam energy spread, we take a flat distribution with a
full width of 1\%~\cite{tor}.

The beamstrahlung calculation requires a set of accelerator
parameters.  For $e^+e^-$, we have used the NLC high-luminosity
parameter set NLC500H~\cite{Nan}.  For $e^-e^-$, we have used the same
parameter set modified for higher $e^-e^-$ luminosity as suggested by
Thompson~\cite{Thompson:2000ij}.  The NLC500H design uses flat beams.
We have also considered an earlier $e^-e^-$ parameter set with round
beams~\cite{Zimmermann:1998bh}, which we call EE500.  In addition, we
have carried out our analysis for the alternative NLC parameter sets
NLC500A,B,C.  These give threshold cross section shapes almost
identical to those with NLC500H.  The luminosities for these designs
are about a factor 3 smaller.

The theoretical cross sections before inclusion of beam effects are
given by the dotted contours in Fig.~\ref{fig:epmcomp}.  In accord
with the angular momentum arguments above, the $\emem$ cross section
rises rapidly at threshold.  In contrast, the $\epem$ cross section
rises extremely slowly.  Of course, these threshold behaviors are
modified after beam effects are included, as seen in the solid
contours.  For flat beams, however, the advantage of $\emem$ beams is
preserved.  For example, 10 GeV above threshold, the $\emem$ cross
section is 990 fb, while the $\epem$ cross section is 2.7 fb.  Note
that the advantage of the $\emem$ mode is compromised for round beams
--- flat $\emem$ beams are essential to preserve the benefits of the
$\beta$ threshold behavior of $\emem \to \ser^- \ser^-$.

The importance of various beam effects on the $\emem$ threshold
behavior is illustrated in Figs.~\ref{fig:expcomp1} and
\ref{fig:expcomp2}.  ISR and beamstrahlung are clearly the dominant
effects, significantly softening the threshold behavior in all cases.
Beam energy spread also smoothes out the threshold behavior, most
noticeably when the selectron width is negligible and the cross
section would rise sharply at threshold otherwise.  Nevertheless, even
after including all beam effects, the $\emem$ cross section rises
rapidly at threshold, and extremely precise measurements are possible,
as we will see below.

\begin{figure}[tbp]
\postscript{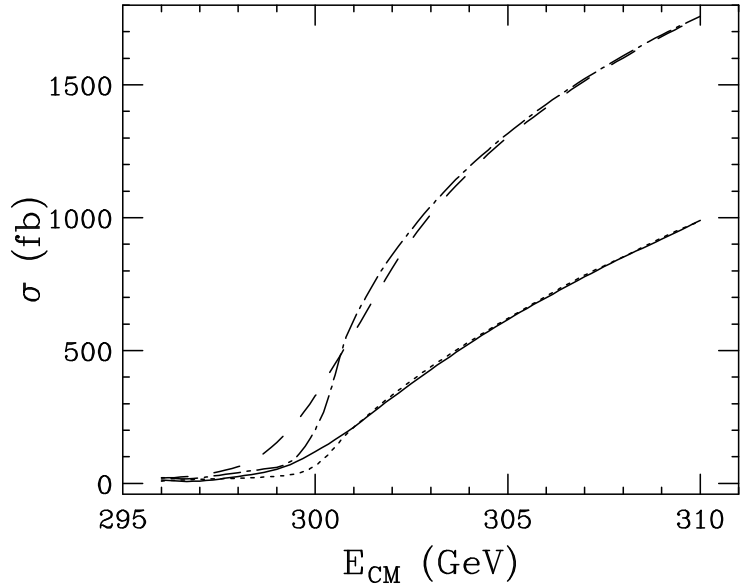}{0.59}
\caption
{Threshold behavior for $\sigma(e^- e^- \to \ser^- \ser^-)$ and
$(\mser, \mchi) = (150~\gev, 100~\gev)$ with no beam effects
(dot-dashed), only ISR/beamstrahlung (dotted), only beam energy spread
(dashed), and both ISR/beamstrahlung and beam energy spread (solid).
$P_{e^-} = 0.8$, and the selectron width of \eqref{width} is assumed.}
\label{fig:expcomp1}
\end{figure}
\begin{figure}[hbtp]
\postscript{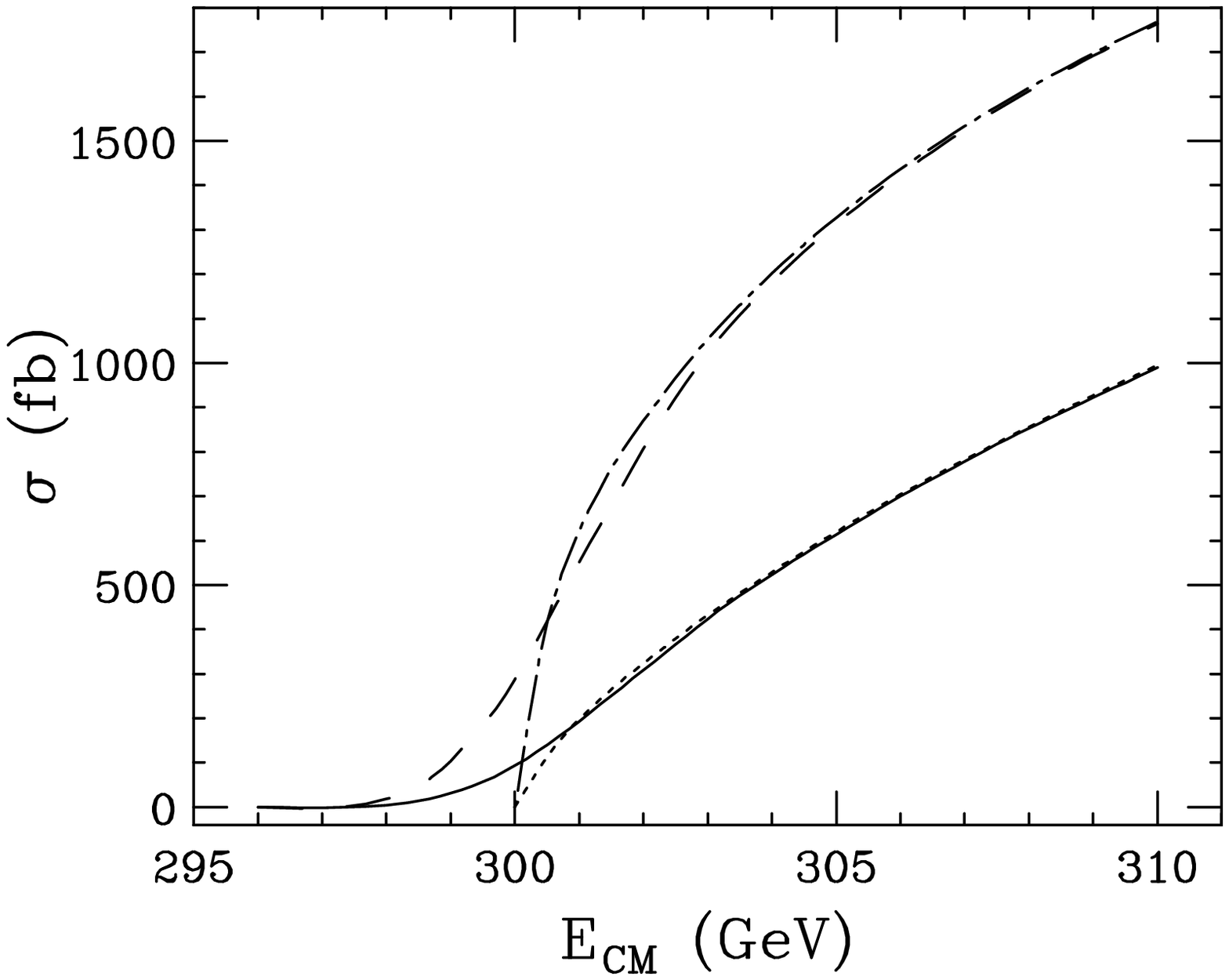}{0.59}
\caption
{As in Fig.~\ref{fig:expcomp1}, but for $\Gamma_{\ser}\approx 0$.}
\label{fig:expcomp2}
\end{figure}

\section{Signal and Backgrounds}
\label{sec:signal}

In supergravity models, which we will focus on here, the selectron
signal is $\emem \to \ser^- \ser^- \to e^- e^- \chi \chi$.  The signal
is two like-sign electrons, each with energy bounded by
\begin{equation}
E_e^{\mathop{}_{\text{\tiny min}}^{\text{\tiny max}} } = 
\frac{1}{2} E_b \left[1 - \frac{\mchi^2}{\mser^2} \right]
\left[ 1 \pm \left( 1 - \frac{\mser^2}{E_b^2} \right)^{1/2} \right] \ ,
\label{range}
\end{equation}
where $E_b$ is the beam energy.  At threshold, the electron spectrum
is mono-energetic. The electrons are emitted isotropically with large
$p_T$ and $\sum p_T \ne 0$.

There are several potential backgrounds to such events, but they may
all be suppressed to negligible levels with little effect on the
signal.  M{\o}ller scattering may be eliminated by a mild acoplanarity
cut.  M{\o}ller scattering with single or double bremsstrahlung may be
eliminated by requiring non-vanishing $\sum p_T$ without visible
photons in the event. $W$ boson pair production, a troublesome
background to selectron pair production in $\epem$ collisions, is
completely eliminated by total lepton number conservation, as is
chargino pair production, even if kinematically allowed.  The two
photon process $\gamma \gamma \to W^+ W^-$, another troublesome
background in $\epem$ collisions, does not produce like-sign
electrons.  The three-body final state $e^-e^-Z$, followed by $Z \to
\nu \bar{\nu}$, is a possible background.  However, the sum of the two
electron energies in these events is greater than $E_b
[1-m_Z^2/(4E_b^2)]$.  For many supersymmetry parameters, including
those considered here, this constraint is inconsistent with
\eqref{range}, and so this background is essentially eliminated by
cuts on the electron energies~\cite{Cuypers:1993vy}.  The background
$e^- \nu W^-$, followed by $W^- \to e^- \bar{\nu}$ may be suppressed
by right-polarizing {\em both\/} beams.  Finally, the four-body
standard model backgrounds $\nu \nu W^- W^-$ and $e^- \nu W^-
Z$~\cite{Cuypers:1994yn} and the three-body supersymmetric
backgrounds, such as $e^- \tilde{\nu} \tilde{W}^-$ and $e^- \tilde{e}
\tilde{B}$, all have cross sections of order 1 fb or less (and in some
cases may also be highly suppressed by beam polarization).

In the end, the dominant background arises from imperfect right-handed
beam polarization leading to $e^- \nu W^-$.  Requiring only that both
electrons have pseudo-rapidity $\eta_{e^-} < 3$ ($5.7^{\circ} <
\theta_{e^-} < 174.3^{\circ}$) and energy $E_{e^-} > 10~\gev$, the
total background is $B \approx 110~\fb \cdot \frac{1}{4} (1-P_{e^-})^2
+ 22~\fb \cdot \frac{1}{2} (1-P_{e^-}^2)$ at center-of-mass energy
$\ecm = 300~\gev$~\cite{Choudhury:1994gm}.  For $P_{e^-} = 0.8$ (0.9),
the background is $B \approx 5.1~\fb$ ($2.4~\fb$). Requiring further
that both electron energies be within the range given by \eqref{range}
will reduce the background to well below the fb level.  The resulting
background is completely negligible in $\emem$ mode, where the signal
cross section quickly rises to hundreds of fb, and cross section
measurements at the 1 fb level are unnecessary for high precision
selectron mass measurements.

In addition to the uncertainty in background under the threshold
signal, there is systematic uncertainty associated with the actual
knowledge of the machine energy calibration.  Not only must the beam
energy be known, but also the differential luminosity spectrum must be
measured to predict the cross section shape in the threshold region.
Fortunately, Wilson has studied these issues in some detail for the
more challenging application of measuring the $W$ mass to 6 MeV with a
scan of the $W^+W^-$ threshold~\cite{WWWilson}.  The beam energy can
be determined to a few MeV with an energy spectrometer, as has been
done at SLC and LEP2.  The differential luminosity spectrum can be
determined from the acollinearity of Bhabha events in the detector
endcaps, and from $e^+e^-\to Z\gamma$ events in which a forward $Z$
decays to leptons.  Scaling down from the $100~\ifb$ proposed by
Wilson to $1~\ifb$, there are still ample statistics in these channels
to reduce the systematic error to much less than 100 MeV.

\section{Mass Determination}
\label{sec:mass}

We now estimate the precision of the selectron mass measurement.  We
consider the case $(\mser, \mchi) = (150~\gev, 100~\gev)$.  The
threshold behavior for these parameters, as well as the 1$\sigma$
statistical error corresponding to $1~\ifb$ at each of seven possible
scan points, is shown in Fig.~\ref{fig:scane-e-mse}.  We assume
$P_{e^-} = 0.8$.  In addition to suppressing background as discussed
in Sec.~\ref{sec:signal}, this beam polarization increases the signal
cross section by $(1+P_{e^-})^2 = 3.24$ relative to the unpolarized
beam case.

The threshold curves for deviations $\Delta\mser = \pm 100~\mev$ from
the central value are also shown in Fig.~\ref{fig:scane-e-mse}.
Clearly, even with $1~\ifb$ of luminosity, deviations in $\mser$ of
order 100 MeV may be distinguished.  Note that for a fixed luminosity
budget, the most stringent constraint on $\mser$ is achieved at $\ecm
\approx 2\mser$.

The identical plot, but for deviations $\Delta \mchi = \pm 10~\gev$,
is given in Fig.~\ref{fig:scane-e-M1}.  For $1~\ifb$, deviations in
$\mchi$ of order 10 GeV are easily distinguished.  For this purpose,
however, measurements at $\ecm \approx 2\mser$ are useless,
and the most incisive constraint is obtained at energies 10 to 20 GeV
above threshold. This is easily understood.  Larger $\mchi$ implies
larger cross sections for $\ecm > 2\mser$, as a result of the Majorana
mass insertion in \eqref{crosssection}, and lower cross sections for
$\ecm < 2\mser$, as a result of the decreased width of \eqref{width}.
These effects cancel at $2\mser$, and so the cross section there is
highly insensitive to $\mchi$.  Note also that, roughly speaking,
deviations in $\mser$ change the normalization of the threshold curve,
while deviations in $\mchi$ change the slope.  These two effects may
therefore be disentangled with data taken at two or more scan points.

\begin{figure}[tbp]
\postscript{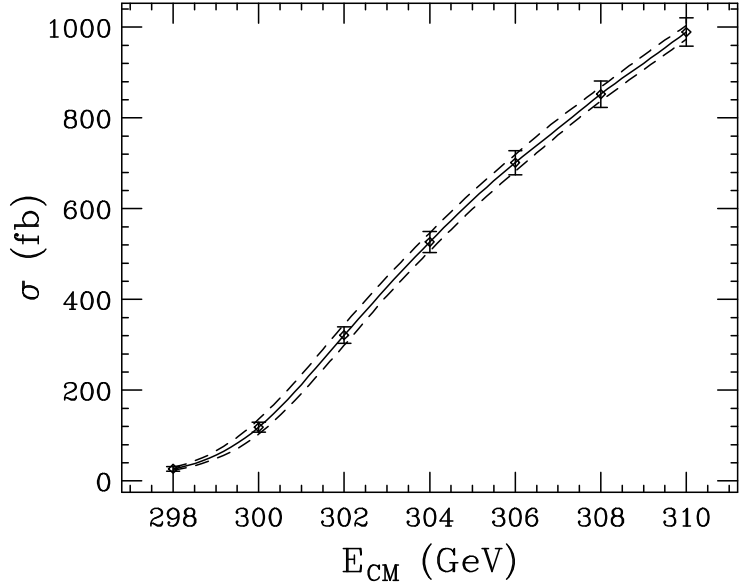}{0.59}
\caption
{Threshold behavior for $\sigma(e^- e^- \to \ser^- \ser^-)$ for
$(\mser, \mchi) = (150~\gev, 100~\gev)$ (solid) and for $\Delta\mser =
\pm 100~\mev$ (dashed). The error bars give the 1$\sigma$ statistical
error corresponding to $1~\ifb$ per point.  $P_{e^-} = 0.8$ , and
ISR/beamstrahlung, beam energy spread, and the selectron width are
included.}
\label{fig:scane-e-mse}
\end{figure}
\begin{figure}[tbp]
\postscript{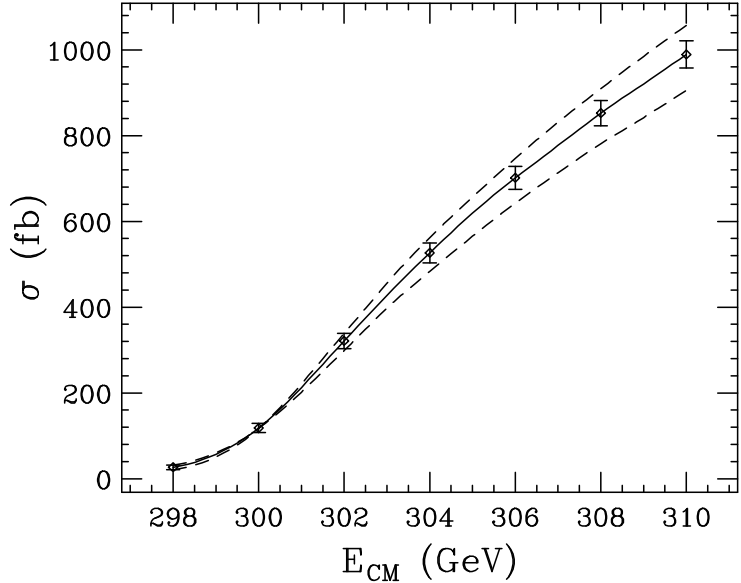}{0.59}
\caption
{Threshold behavior for $\sigma(e^- e^- \to \ser^- \ser^-)$ for
$(\mser, \mchi) = (150~\gev, 100~\gev)$ (solid) and for $\Delta\mchi =
\pm 10~\gev$ (dashed). The error bars give the 1$\sigma$ statistical
error corresponding to $1~\ifb$ per point.  $P_{e^-} = 0.8$, and
ISR/beamstrahlung, beam energy spread, and the selectron width are
included.}
\label{fig:scane-e-M1}
\end{figure}

To determine the precision with which $\mser$ and $M_1$ may be
constrained in a threshold scan, we use the binned likelihood method.
We define
\begin{equation}
\ln {\cal L}(\mser, M_1) \equiv \sum_i N'_i \ln N_i
(\mser, M_1) - N_i (\mser, M_1) \ ,
\end{equation}
where the sum is over scan points. $N'_i$ is the measured number of
events at scan point $i$, which we take to be the theoretical
prediction given the underlying physical parameters, and $N_i (\mser,
M_1)$ is the predicted number of events given hypothetical parameters
$\mser$ and $M_1$.  The parameter $\ln {\cal L}$ is maximized for the
true underlying values of the parameters, and the width of the $\ln
{\cal L}$ peak determines the precision with which these parameters
are measured, with $\chi^2 \equiv 2 (\ln {\cal L}_{\text{max}} - \ln
{\cal L})$ the squared standard deviation.

The optimal scan strategy depends crucially on what information is
known beforehand from other processes and which parameter one most
hopes to constrain. These are complicated issues.  Here we consider
two possibilities.  First, to constrain both parameters, one might
split the luminosity evenly between $\ecm = 2\mser$ and $2\mser +
10~\gev$ in a `2-point scan.'  $\chi^2$ contours in the $(\mser, M_1)$
plane are given in Fig.~\ref{fig:mse_M1_2pt}.  For a total integrated
luminosity $\ltot = 10~\ifb$, the 90\% C.L. ($\chi^2 = 4.61$)
ellipse (not shown) is bounded by $\mser = 150 \pm 0.065~\gev$ and
$M_1 = 100 ^{+5}_{-4}~\gev$. The neutralino mass is poorly constrained
this way, and is likely to be determined more precisely through
kinematic endpoints.  In this case, projecting the $\chi^2=1$ ellipse
down to the $\mser$ axis gives
\begin{equation}
\text{2-point scan:}\ 
\ltot^{\emem} = 1\ (10)~\ifb \Longrightarrow
\Delta\mser = 90\ (30)~\mev \ (1\sigma)\ .
\end{equation}

On the other hand, given that the neutralino mass is likely to be
better measured by other methods, one might simple desire to constrain
the selectron mass.  The optimal strategy is then to concentrate all
of the luminosity at $\ecm = 2\mser$, where the sensitivity to $\mser$
is greatest.  Results of this `$\mser$-optimized scan' are given in
Fig.~\ref{fig:mse_M1_1pt}.  As expected, the neutralino mass is
completely unconstrained.  However, given some modest constraints on
the neutralino mass from some other source, we find
\begin{equation}
\text{$\mser$-optimized scan:}\ 
\ltot^{\emem} = 1\ (10)~\ifb \Longrightarrow
\Delta\mser = 70\ (20)~\mev \ (1\sigma)\ .
\end{equation}
Selectron mass measurements below the part per mil level are therefore
possible with meager investments of luminosity.

\begin{figure}[tbp]
\postscript{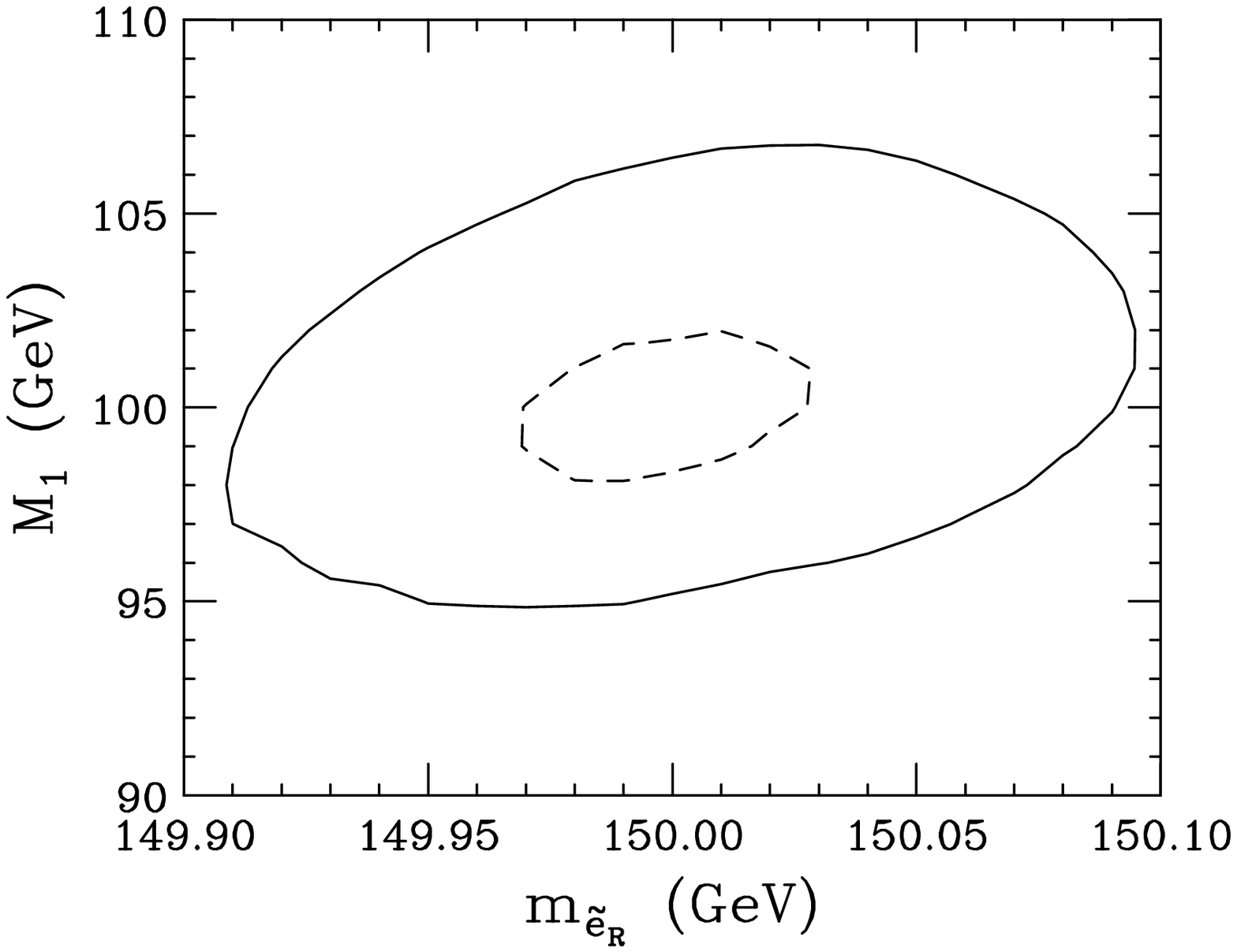}{0.59}
\caption
{$\chi^2 = 1$ constraint contours in the $(\mser, M_1)$ plane for the
`2-point scan' of $\sigma(e^- e^- \to \ser^- \ser^-)$ for $\ltot =
1~\ifb$ (solid) and $10~\ifb$ (dashed).  The luminosity is divided
equally between $\ecm = 300~\gev$ and 310 GeV.  $P_{e^-} = 0.8$, and
ISR/beamstrahlung, beam energy spread, and the selectron width of
\eqref{width} are included.}
\label{fig:mse_M1_2pt}
\end{figure}
\begin{figure}[tbp]
\postscript{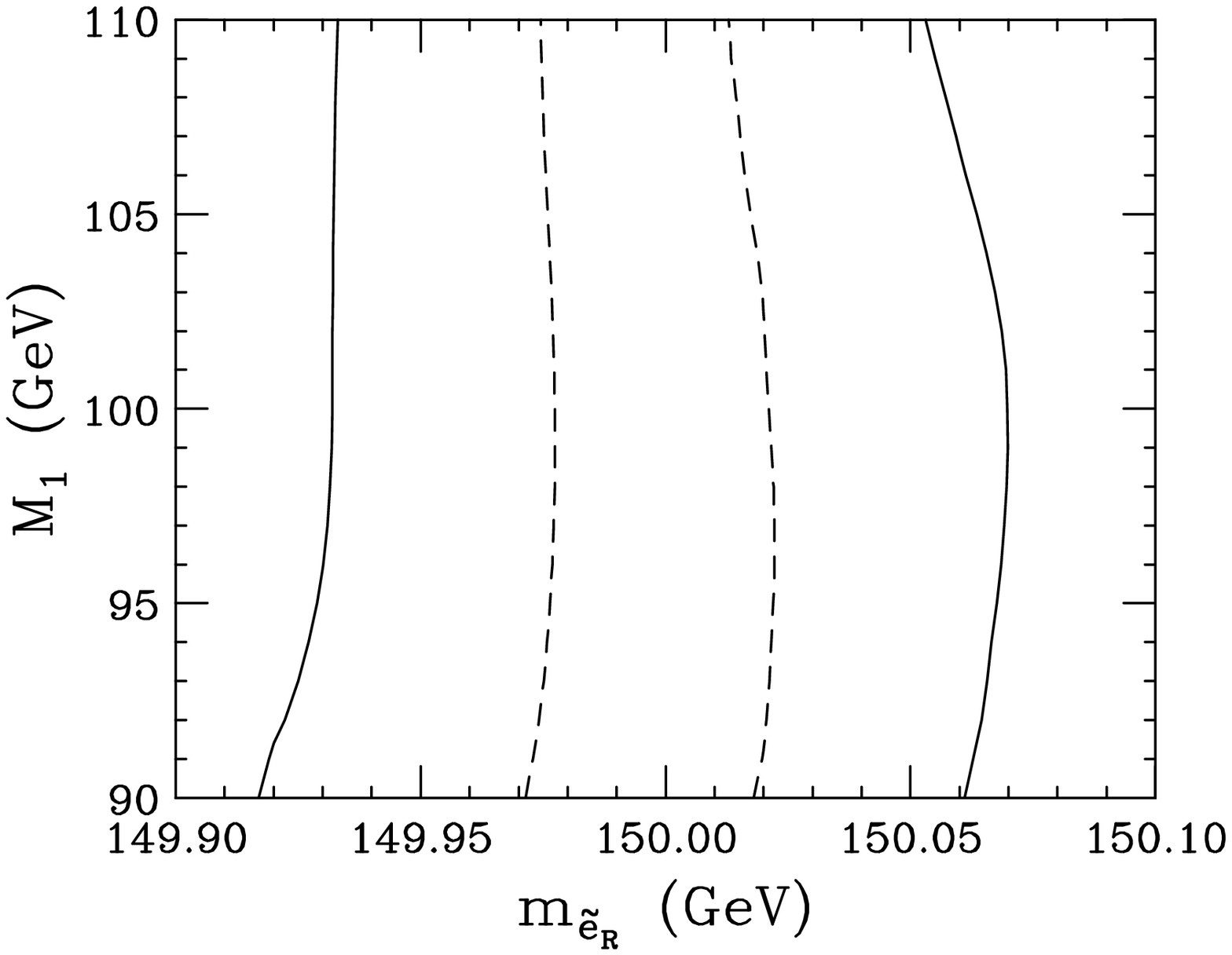}{0.59}
\caption
{$\chi^2 = 1$ constraint contours in the $(\mser, M_1)$ plane for the
`$\mser$-optimized scan' of $\sigma(e^- e^- \to \ser^- \ser^-)$ for
$\ltot = 1~\ifb$ (solid) and $10~\ifb$ (dashed).  The luminosity is
concentrated at $\ecm = 300~\gev$.  $P_{e^-} = 0.8$, and
ISR/beamstrahlung, beam energy spread, and the selectron width of
\eqref{width} are included.}
\label{fig:mse_M1_1pt}
\end{figure}

\section{Comparison with $\bold{\epem}$ mode}
\label{sec:comparison}

We now compare the results of the previous section with what can be
achieved in $\epem$ collisions.  The cross section $\sigma(\epem \to
\ser^+ \ser^-)$ rises as $\beta^3$ at threshold.  Values of ${\cal
O}(1)$ fb are therefore typical even $\sim 10~\gev$ above
threshold. In addition, backgrounds such as $\epem \to W^+W^-, e^- \nu
W^+$ and $\gamma \gamma \to W^+ W^-$ are large and difficult to
eliminate.  This contrasts sharply with the $\emem$ case, where the
signal is large, and the analogues of these backgrounds are absent or
easily suppressed. Detailed studies of these and other backgrounds, as
well as the cuts required to remove them, are necessary to fully
understand the potential of $\epem$ threshold studies.  In this
section we make the most optimistic assumption possible, namely, we
neglect all backgrounds.  Our conclusion that very large luminosities
are required in $\epem$ collisions will only be strengthened with more
detailed analyses.

In Figs.~\ref{fig:scane+e-mse} and \ref{fig:scane+e-M1} we present
threshold cross sections for $\sigma(e^+ e^- \to \ser^+ \ser^-)$ for
$(\mser, \mchi) = (150~\gev, 100~\gev)$, as well as for deviations in
$\mser$ and $\mchi$.  The cross sections are small, and the
statistical error bars shown are for $100~\ifb$ per point, in contrast
to the $1~\ifb$ assumed in Figs.~\ref{fig:scane-e-mse} and
\ref{fig:scane-e-M1}.  Note also that, in contrast to the $\emem$
case, deviations in $\mser$ and $M_1$ have the same qualitative effect
on the threshold curve --- roughly speaking, both change the
normalization.  The effect of increasing $\mser$ is therefore nearly
indistinguishable from the effect of decreasing $M_1$, and the
degeneracy is difficult to remove by threshold scans alone.

\begin{figure}[tbp]
\postscript{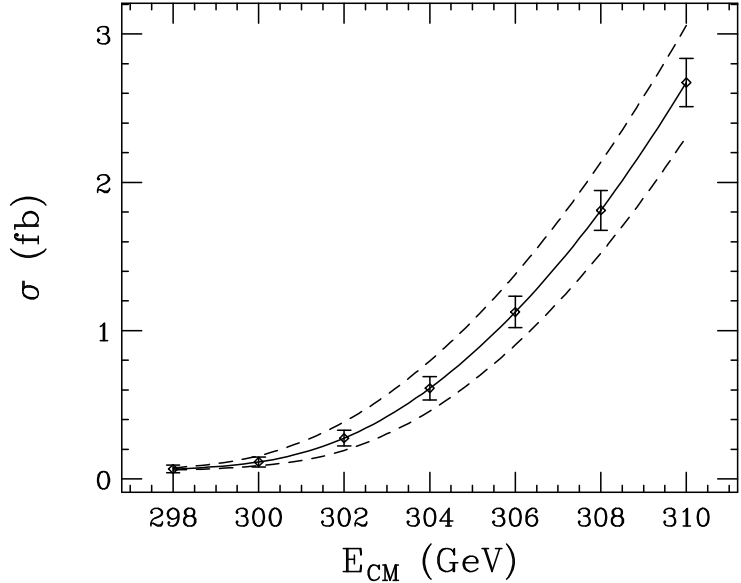}{0.59}
\caption
{Threshold behavior for $\sigma(e^+ e^- \to \ser^+ \ser^-)$ for
$(\mser, \mchi) = (150~\gev, 100~\gev)$ (solid) and for $\Delta\mser =
\pm 400~\mev$ (dashed). The error bars give the 1$\sigma$ statistical
error corresponding to $100~\ifb$ per point. $P_{e^-} = 0.8$, $P_{e^+}
= 0$, and ISR/beamstrahlung, beam energy spread, and the selectron
width are included.}
\label{fig:scane+e-mse}
\end{figure}
\begin{figure}[hbtp]
\postscript{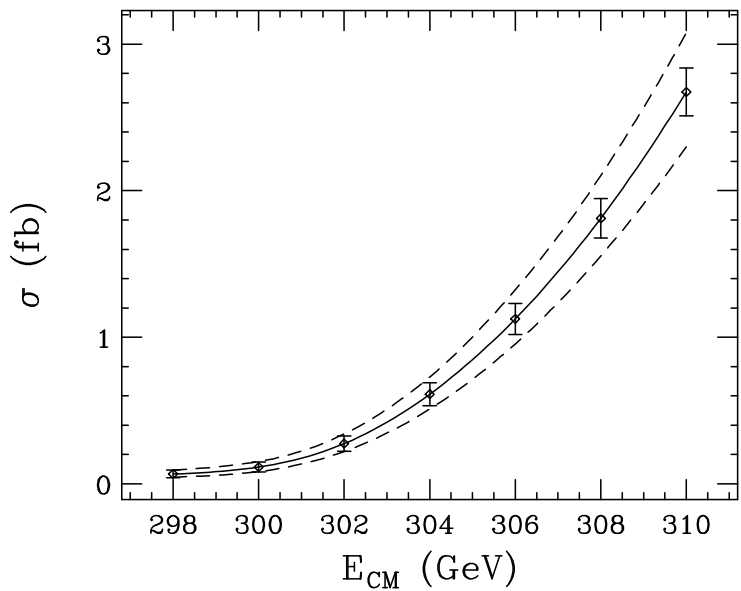}{0.59}
\caption
{Threshold behavior for $\sigma(e^+ e^- \to \ser^+ \ser^-)$ for
$(\mser, \mchi) = (150~\gev, 100~\gev)$ (solid) and for $\Delta\mchi =
\pm 4~\gev$ (dashed). The error bars give the 1$\sigma$ statistical
error corresponding to $100~\ifb$ per point. $P_{e^-} = 0.8$, $P_{e^+}
= 0$, and ISR/beamstrahlung, beam energy spread, and the selectron
width are included.}
\label{fig:scane+e-M1}
\end{figure}

As evident in Figs.~\ref{fig:scane+e-mse} and \ref{fig:scane+e-M1},
data taken at any of the potential scan points provides roughly the
same information.  We consider a 2-point scan with luminosity divided
equally between $\ecm = 300~\gev$ and 310 GeV; results vary little for
different scan strategies.  The $\chi^2$ contours are given in
Fig.~\ref{fig:mse_M1_e+e-}.  As expected, from threshold data it is
very difficult to determine $\mser$ and $M_1$ separately.  In contrast
to the $\emem$ case, one must necessarily rely on kinematic endpoints
to break this degeneracy.  Assuming the Bino mass is known {\em
exactly}, we find
\begin{equation}
\text{2-point scan:}\ 
\ltot^{\epem} = 100\ (1000)~\ifb \Longrightarrow
\Delta\mser = 210\ (70)~\mev \ (1\sigma)\ .
\end{equation}
If the Bino mass is known only to 1 GeV, these bounds become
$\ltot^{\epem} = 100\ (1000)~\ifb \Longrightarrow \Delta\mser = 290\
(140)~\mev \ (1\sigma)$.

\begin{figure}[tbp]
\postscript{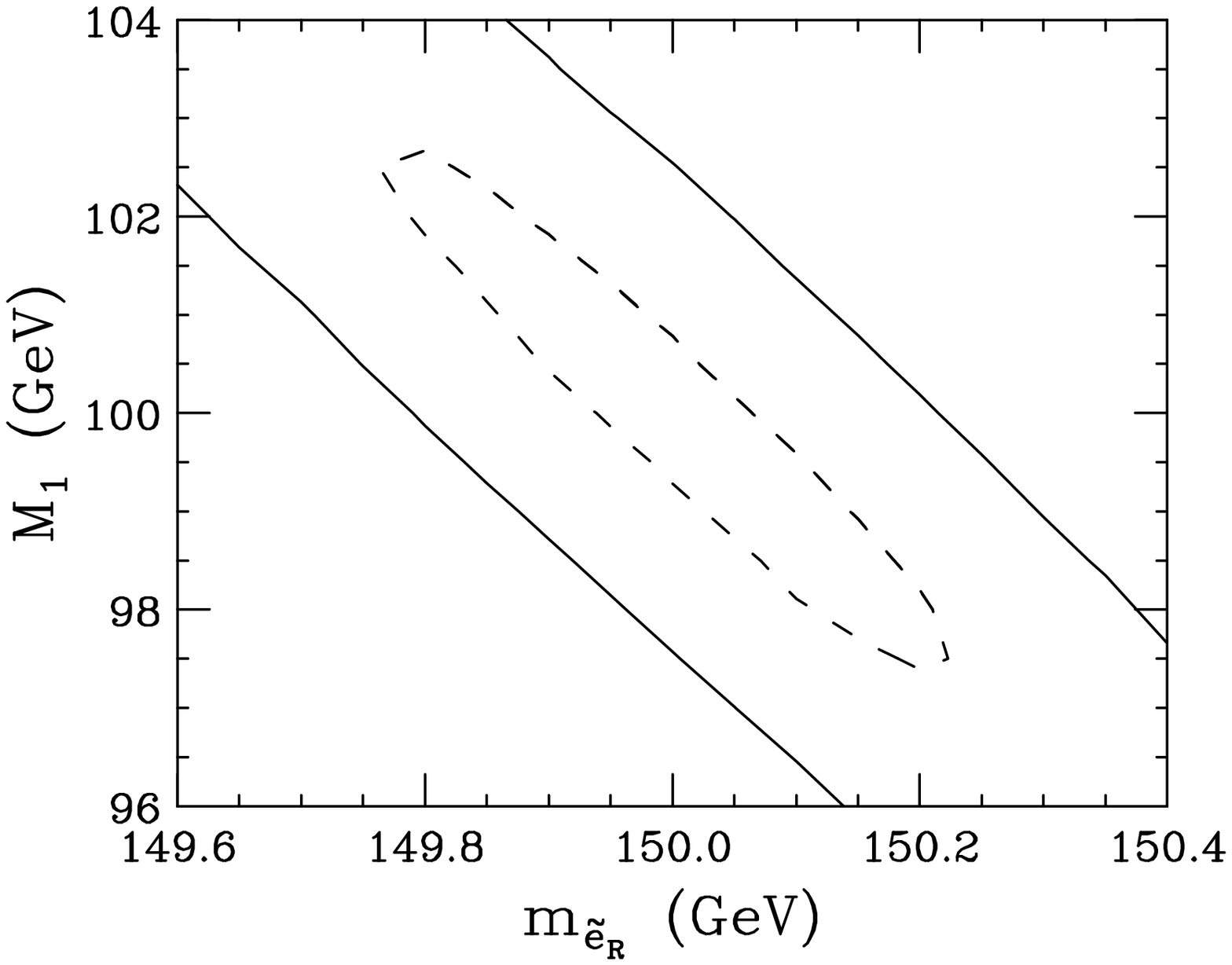}{0.59}
\caption
{$\chi^2 = 1$ constraint contours in the $(\mser, M_1)$ plane for the
`2-point scan' of $\sigma(\epem \to \ser^+ \ser^-)$ for $\ltot =
100~\ifb$ (solid) and $1000~\ifb$ (dashed).  The luminosity is divided
equally between the points $\ecm = 300~\gev$ and 310 GeV.  $P_{e^-} =
0.8$, $P_{e^+} = 0$, and ISR/beamstrahlung, beam energy spread, and
the selectron width are included.}
\label{fig:mse_M1_e+e-}
\end{figure}

Threshold scans in $\epem$ colliders have been studied previously in
Refs.~\cite{Martyn:1999tc,Martyn:1999xc}, where measurements of a wide
variety of superparticle masses were considered.  While our results
agree qualitatively, we are unable to reproduce their results in
detail.  In Ref.~\cite{Martyn:1999tc} the authors consider the
scenario $(\mser, \mchi) = (132~\gev, 71.9~\gev)$.  Assuming $P_{e^-}
= 0.8$, $P_{e^+} = 0.6$, and $\ltot^{\epem} = 100~\ifb$ divided
equally between the ten points $\ecm = 265, 266, \ldots, 274~\gev$,
they found $\Delta \mser = 50~\mev$. With the same assumptions, we
find $\Delta \mser = 90~\mev$ ($1\sigma$) if $M_1$ is known exactly,
and $\Delta \mser = 130~\mev$ ($1\sigma$) if $M_1$ is known to 1 GeV.
Our bounds are significantly less stringent --- to achieve $\Delta
\mser = 50~\mev$, we find that at least $\ltot^{\epem} = 320~\ifb$ is
required. We stress again that in both analyses, backgrounds are
neglected.  Once included, the achievable precisions in $\epem$
colliders will certainly deteriorate, possibly significantly.

\section{Summary and Discussion}
\label{sec:summary}

The $\emem$ mode is an inexpensive and simple extension of the linear
collider program.  We have described an important virtue of this mode
for studies of supersymmetry, namely, the measurement of selectron
mass at threshold.  In $\emem$ mode, many potential backgrounds to
selectron pair production are simply absent, and those that remain may
be suppressed to negligible levels with double beam polarization.  In
addition, the unique quantum numbers of the $\emem$ initial state lead
to large cross sections even slightly above threshold, in contrast to
the case of $\epem$ colliders.  We have included the ISR/beamstrahlung
and beam energy spread of realistic beam designs and find that
selectron mass measurements below 100 MeV level are possible with only
$\ltot = 1~\ifb$, or less than a week of running at design luminosity.
In $\epem$ collisions, such precision, even ignoring large
backgrounds, requires more than two orders of magnitude more
luminosity.

Throughout this study, we have assumed that the lightest neutralino is
a pure Bino, and that slepton flavor violation is absent.  It is, of
course, important that these assumptions be verifiable experimentally.
Note that the results derived here are not dependent on extremely
precise cross section measurements.  The statistical uncertainties at
individual scan points are typically of order 10\%, and so the impact
of Bino purity and other complications need only be constrained to be
below this level.

The neutralino mixing matrix may be constrained most directly by
discovering all four neutralinos and two charginos.  If they are
within kinematic reach of a linear collider, discovery is guaranteed,
and their masses and other observables will allow a highly accurate
determination of the neutralino mass matrix.  Alternatively, if some
states, such as the Higgsinos, are beyond reach, observables such as
$\sigma(e^+ e^-_R \to \chi^+ \chi^-)$~\cite{Feng:1995zd} may be used
to reduce the theoretical uncertainty in $\sigma(\emem \to \ser^-
\ser^-)$ to sufficient levels.  Slepton flavor violation may also
change the prediction for $\sigma(\emem \to e^- e^- \chi \chi)$.
However, the resulting signals, such as $\sigma(\emem \to e^- \mu^-
\chi \chi)$ are so spectacular that they will be stringently bounded,
or, if seen, precisely measured~\cite{Arkani-Hamed:1996au}. Such
effects, then, will not lead to large theoretical uncertainties.
Finally, of course, at loop-level, many unknown supersymmetry
parameters enter.  However, these are unlikely to disrupt the
theoretical calculations of threshold cross sections at the 10\%
level.

The study described here is but one use of the peculiar features of
the $\emem \to \ser^- \ser^-$ reaction.  If the lightest
supersymmetric particle is Higgsino-like, or in theories with
$R$-parity violation or low-energy supersymmetry breaking, the Bino
mass parameter $M_1$ may be very large.  As a result of the Bino mass
insertion in Fig.~\ref{fig:selectron}, the cross section for $\emem
\to \ser^- \ser^-$ is large even for large $M_1$, and a high precision
measurement of $M_1$ is possible even for $M_1 \sim
1~\tev$~\cite{Feng:1998ud,Blochinger:2000kp}.  In addition, the full
arsenal of linear collider modes may allow one to extend the high
precision measurement of $\mser$ to the rest of the first generation
sleptons through a series of $\beta$ threshold scans: $e^-e^- \to
\tilde{e}^-_R \tilde{e}^-_R$ yields $m_{\tilde{e}_R}$; $e^+e^- \to
\tilde{e}^{\pm}_R \tilde{e}^{\mp}_L$ yields $m_{\tilde{e}_L}$; $e^+e^-
\to \chi^+ \chi^-$ yields $m_{\chi^{\pm}}$; and $e^- \gamma \to
\tilde{\nu}_e \chi^-$ yields $m_{\tilde{\nu}_e}$~\cite{Barger:1998qu}.
The quantity $\msel - \mser$ gives a highly model-independent
measurement of $\tb$~\cite{Feng:1998ud}.  More generally, as noted
previously, precise measurements of all of these masses will play an
essential role in the program of extrapolating weak scale parameters
to higher energy scales to uncover a more fundamental theory of
nature.

\section*{Acknowledgments}

We thank Claus Bl\"ochinger and Kathy Thompson for helpful
conversations and Clemens Heusch for encouragement.  The work of JLF
was supported in part by the U.~S.~Department of Energy under
cooperative research agreement DF--FC02--94ER40818.  The work of MEP
was supported by the Department of Energy, contract
DE--AC03--76SF00515.


\begin{thebibliography}{99}

\bibitem{Murayama:1996ec}
H.~Murayama and M.~E.~Peskin,
%``Physics opportunities of e+ e- linear colliders,''
Ann.\ Rev.\ Nucl.\ Part.\ Sci.\ {\bf 46}, 533 (1996)
[hep-ex/9606003].
%%CITATION = HEP-EX 9606003;%%

\bibitem{95} 
{\em Proceedings of the 1st International Workshop on
Electron-Electron Interactions at TeV Energies ($e^-e^-$95)}, Santa
Cruz, California, 5--6 September 1995, ed. C.~A.~Heusch,
Int.\ J.\ Mod.\ Phys.\ A {\bf 11}, 1523-1697 (1996).

\bibitem{97} 
{\em Proceedings of the 2nd International Workshop on
Electron-Electron Interactions at TeV Energies ($e^-e^-$97)}, Santa
Cruz, California, 22--24 September 1997, ed. C.~A.~Heusch,
Int.\ J.\ Mod.\ Phys.\ A {\bf 13}, 2217-2549 (1998).

\bibitem{99} 
{\em Proceedings of the 3nd International Workshop on
Electron-Electron Interactions at TeV Energies ($e^-e^-$99)}, Santa
Cruz, California, 10--12 December 1999, ed. C.~A.~Heusch,
Int.\ J.\ Mod.\ Phys.\ A {\bf 15}, 2347-2628 (2000).

\bibitem{Keung:1983nq}
W.~Y.~Keung and L.~Littenberg,
%``Test Of Supersymmetry In E- E- Collision,''
Phys.\ Rev.\ D {\bf 28}, 1067 (1983).
%%CITATION = PHRVA,D28,1067;%%

\bibitem{Feng:1998ud}
J.~L.~Feng,
%``Supersymmetry at linear colliders: The importance of 
%being e- e-,''
Int.\ J.\ Mod.\ Phys.\ A {\bf 13}, 2319 (1998)
[hep-ph/9803319].
%%CITATION = HEP-PH 9803319;%%

\bibitem{Feng:2000zv}
J.~L.~Feng,
%``Physics at e- e- colliders,''
Int.\ J.\ Mod.\ Phys.\ A {\bf 15}, 2355 (2000)
[hep-ph/0002055].
%%CITATION = HEP-PH 0002055;%%

\bibitem{Arkani-Hamed:1996au}
N.~Arkani-Hamed, H.~Cheng, J.~L.~Feng and L.~J.~Hall,
%``Probing Lepton Flavor Violation at Future Colliders,''
Phys.\ Rev.\ Lett.\ {\bf 77}, 1937 (1996)
[hep-ph/9603431].
%%CITATION = HEP-PH 9603431;%%

\bibitem{Arkani-Hamed:1997km}
N.~Arkani-Hamed, J.~L.~Feng, L.~J.~Hall and H.~Cheng,
%``CP violation from slepton oscillations at the LHC and NLC,''
Nucl.\ Phys.\ B {\bf 505}, 3 (1997)
[hep-ph/9704205].
%%CITATION = HEP-PH 9704205;%%

\bibitem{Cheng:1997sq}
H.~Cheng, J.~L.~Feng and N.~Polonsky,
%``Super-oblique corrections and non-decoupling of supersymmetry
%breaking,''
Phys.\ Rev.\ D {\bf 56}, 6875 (1997)
[hep-ph/9706438];
%%CITATION = HEP-PH 9706438;%%
%\bibitem{Cheng:1998vy}
%H.~Cheng, J.~L.~Feng and N.~Polonsky,
%``Signatures of multi-TeV scale particles in supersymmetric
%theories,''
Phys.\ Rev.\ D {\bf 57}, 152 (1998)
[hep-ph/9706476].
%%CITATION = HEP-PH 9706476;%%

\bibitem{Cheng:1998sn}
H.~Cheng,
%``Precision supersymmetry measurements at the e- e- collider,''
Int.\ J.\ Mod.\ Phys.\ A {\bf 13}, 2329 (1998)
[hep-ph/9801234].
%%CITATION = HEP-PH 9801234;%%

\bibitem{Blair:2001gy}
G.~A.~Blair, W.~Porod and P.~M.~Zerwas,
%``Reconstructing supersymmetric theories at high energy scales,''
Phys.\ Rev.\ D {\bf 63}, 017703 (2001)
[hep-ph/0007107].
%%CITATION = HEP-PH 0007107;%%

\bibitem{PANDORA}
M.~E.~Peskin, {\tt pandora 2.1}, 
http://www-sldnt.slac.stanford.edu/nld/new/Docs/ \\
Generators/PANDORA.htm.

\bibitem{Martyn:1999tc}
H.~Martyn and G.~A.~Blair,
%``Determination of sparticle masses and SUSY parameters,''
hep-ph/9910416.
%%CITATION = HEP-PH 9910416;%%

\bibitem{Martyn:1999xc}
H.~U.~Martyn,
%``Width effects in slepton production e+ e- --> smuon(R)+
%smuon(R)-,''
hep-ph/0002290.
%%CITATION = HEP-PH 0002290;%%

\bibitem{Peskin:1998jy}
M.~E.~Peskin,
%``Systematics of slepton production in e+ e- and e- e- collisions,''
Int.\ J.\ Mod.\ Phys.\ A {\bf 13}, 2299 (1998)
[hep-ph/9803279].
%%CITATION = HEP-PH 9803279;%%

\bibitem{Skrzypek:1991qs}
M.~Skrzypek and S.~Jadach,
%``Exact and approximate solutions for the electron nonsinglet 
%structure function in QED,''
Z.\ Phys.\ C {\bf 49}, 577 (1991).
%%CITATION = ZEPYA,C49,577;%%

\bibitem{myLC}
M.~E.~Peskin, Linear Collider Collaboration technical note LCC--0010
(1999).

\bibitem{YC}
K.~Yokoya and P.~Chen, in {\em Proceedings of the 1989 Particle
Accelerator Conference}, eds. F.~Bennett and L.~Taylor (IEEE, 1989).

\bibitem{Chen1}
P.~Chen,
%``Differential luminosity under multi - photon beamstrahlung,''
Phys.\ Rev.\ D {\bf 46}, 1186 (1992).
%%CITATION = PHRVA,D46,1186;%%

\bibitem{Chen2}
P.~Chen, T.~L.~Barklow and M.~E.~Peskin,
%``Hadron production in gamma gamma collisions as a background for 
%e+ e- linear colliders,''
Phys.\ Rev.\ D {\bf 49}, 3209 (1994)
[hep-ph/9305247].
%%CITATION = HEP-PH 9305247;%%

\bibitem{Chen3}
K.~A.~Thompson and P.~Chen,
%``Luminosity and disruption in e- e- linear colliders,''
in {\em Physics and Experiments at Future Linear $e^+e^-$ Colliders}
(LCWS99), eds. E. Fernandez and A. Pacheco (U. Auton. de Barcelona,
Bellaterra, 2000).

\bibitem{Thompson:2000ij}
K.~A.~Thompson,
%``Optimization of NLC luminosity for e- e- running,''
Int.\ J.\ Mod.\ Phys.\ A {\bf 15}, 2485 (2000).
%%CITATION = IMPAE,A15,2485;%%

\bibitem{tor}
T.~Raubenheimer, private communication.

\bibitem{Nan}
T.~Raubenheimer, talk presented at the 5th International Linear
Collider Workshop, Fermilab, 24-28 October 2000.

\bibitem{Zimmermann:1998bh}
F.~Zimmermann, K.~A.~Thompson and R.~H.~Helm,
%``Electron electron luminosity in the next linear collider: 
%A preliminary study,''
Int.\ J.\ Mod.\ Phys.\ A {\bf 13}, 2443 (1998).
%%CITATION = IMPAE,A13,2443;%%

\bibitem{Cuypers:1993vy}
F.~Cuypers, G.~J.~van Oldenborgh and R.~R\"uckl,
%``Supersymmetric signals in e- e- collisions,''
Nucl.\ Phys.\ B {\bf 409}, 128 (1993)
[hep-ph/9305287].
%%CITATION = HEP-PH 9305287;%%

\bibitem{Cuypers:1994yn}
F.~Cuypers, K.~Kolodziej and R.~R\"uckl,
%``Standard model predictions for weak boson pair production in e- e-
%scattering,''
Nucl.\ Phys.\ B {\bf 430}, 231 (1994)
[hep-ph/9405421].
%%CITATION = HEP-PH 9405421;%%

\bibitem{Choudhury:1994gm}
D.~Choudhury and F.~Cuypers,
%``Electron-electron scattering as a probe of anomalous gauge
%couplings,''
Nucl.\ Phys.\ B {\bf 429}, 33 (1994)
[hep-ph/9405212].
%%CITATION = HEP-PH 9405212;%%

\bibitem{WWWilson}
G.~W.~Wilson, 
in {\em Physics and Experiments at Future Linear $e^+e^-$ Colliders}
(LCWS99), eds. E.~Fernandez and A.~Pacheco (U. Auton. de Barcelona,
Bellaterra, 2000).

\bibitem{Feng:1995zd}
J.~L.~Feng, M.~E.~Peskin, H.~Murayama and X.~Tata,
%``Testing supersymmetry at the next linear collider,''
Phys.\ Rev.\ D {\bf 52}, 1418 (1995)
[hep-ph/9502260].
%%CITATION = HEP-PH 9502260;%%}

\bibitem{Blochinger:2000kp}
C.~Bl\"ochinger, H.~Fraas, T.~Mayer and G.~Moortgat-Pick,
%``Determination of the gaugino mass parameter M(1) in different
%linear collider modes,''
hep-ph/0101176.
%%CITATION = HEP-PH 0101176;%%

\bibitem{Barger:1998qu}
V.~Barger, T.~Han and J.~Kelly,
%``Sparticle production in electron photon collisions,''
Phys.\ Lett.\ B {\bf 419}, 233 (1998)
[hep-ph/9709366].
%%CITATION = HEP-PH 9709366;%%

\end{thebibliography}
\end{document}